\documentclass[prx,twocolumn]{revtex4-1}

\usepackage[utf8]{inputenc}
\usepackage{graphicx}

\usepackage{amsmath}
\usepackage{amssymb}

\newcommand{\ket}[1]{\left| #1 \right>}

\newcommand{\ketbra}[2]{\left| #1 \middle> \middle< #2 \right|}

\begin{document}
\title{Optomechanical analogy for toy cosmology with quantized scale factor}
\author{Joseph A. Smiga$^{1,2}$ and J. M. Taylor$^{1,2,3}$}
\affiliation{$^1$Joint Center for Quantum Information and Computer Science, College Park, MD, USA}
\affiliation{$^2$Joint Quantum Institute/NIST, College Park, MD, USA}
\affiliation{$^3$Research Center for Advanced Science and Technology, Tokyo, Japan}

\begin{abstract}

The simplest cosmology --- the Friedmann-Robertson-Walker-Lema\^{i}tre (FRW) model --- describes a spatially homogeneous and isotropic universe where the scale factor is the only dynamical parameter. Here we consider how quantized electromagnetic fields become entangled with the scale factor in a toy version of the FRW model. A system consisting of a photon, source, and detector is described in such a universe, and we find that the detection of a redshifted photon by the detector system constrains possible scale factor superpositions. Thus, measuring the redshift of the photon is equivalent to a weak measurement of the underlying cosmology. We also consider a potential optomechanical analogy system that would enable experimental exploration of these concepts. The analogy focuses on the effects of photon redshift measurement as a quantum back-action on metric variables, where the position of a movable mirror plays the role of the scale factor. By working in the rotating frame, an effective Hubble equation can be simulated with a simple free moving mirror.
\end{abstract}
\maketitle

\section{Introduction}

Quantum mechanics provides several key mechanisms for proposed explanations of cosmological structure~\cite{bojowald_quantum_2015}. 
For example, one
of the major observational discoveries in cosmology is that the universe is expanding~\cite{hubble_relation_1929,sitter_distances_1930}. Many puzzles about this expansion, 
particularly at early time, seem to hinge upon the quantum behavior of the universe's structure, that is, its metric~\cite{bardeen_spontaneous_1983}. 
A key question in these puzzles is the role decoherence may play in freezing-in quantum fluctuations~\cite{kibble_implications_1980}. 
In establishing a model for the universe, one would expect macroscopic objects/events to behave classically. Equivalently, one would expect macroscopic objects to be highly entangled such that the state of an object can be easily restricted by making any related measurement. Thus, any quantum treatment of cosmological variables should result in states highly entangled with that variable.

Some simple models consider various toy cosmologies in which parameters of the metric are treated as quantum variables. For example, ref.~\cite{ozer_implications_1998} assumes that the universe's metric is described by a position wavefunction and considers the resulting implications. Another approach is to describe the cosmology using quantum Brownian motion in a parametric bath as in ref.~\cite{hu_quantum_1994,calzetta_noise_1994}. Here we consider a toy cosmology based upon the Friedmann-Robertson-Walker-Lema\^{i}tre (FRW) model, where the scale factor is quantized, and a wavefunction describing the state of the universe is defined. This wavefunction is often explored in the context of its dynamics as constrained by the Wheeler-DeWitt equation~\cite{alvarenga_troubles_2003,ren_hamiltonian_2007,majumder_quantum_2011,bouhmadi-lopez_frw_2005,khvedelidze_generalized_2001}. Recently, there has been some suggestion that weak measurement of metric variables might act analogously to dark energy, again in this quantized FRW setting~\cite{altamirano_emergent_2016}. 

In this paper, the properties of a simple but quantum metric are explored to help us illustrate the role weak measurements --- here mediated by photon emission and absorption --- play in stabilizing the metric towards a `classical' state. Thus, all of the underlying factors driving cosmological dynamics (e.g., a perfect fluid~\cite{alvarenga_troubles_2003}, imperfect fluid~\cite{ren_hamiltonian_2007}, Chaplygin gas~\cite{majumder_quantum_2011,bouhmadi-lopez_frw_2005}, or field~\cite{khvedelidze_generalized_2001}) are ignored, and we do not assume any constraints on dynamics (e.g., the Wheeler-DeWitt equation). Instead, the dynamics of the universe are essentially assumed.
Working in conformal time, we show how photon emission and later absorption leads to a strong entanglement between sources, detectors, and the metric variables. This provides a tool for examining the intuitive result that redshifted photons convey information about the universe's past and effectively decohere potential superpositions of metric variables. 

Direct cosmological experiments are highly limited due to the space and time scales associated with the dynamics. Thus, various analogies are experimentally implemented to explore cosmological effects. For example, superfluid helium models are used to describe freezing-in quantum fluctuations~\cite{zurek_cosmological_1985,zurek_cosmological_1996}. In addition, experiments with Bose-Einstein condensates (BEC) have demonstrated an analogy to Hawking radiation found with black holes~\cite{steinhauer_observation_2016}. Other analogies consider emulating the space-time, such as with BEC~\cite{matos_space-time_2015}, graphene impurities~\cite{cortijo_electronic_2007}, and in an optical lattice~\cite{szpak_curved_2014}. Using the toy FRW model detailed in this paper, we describe an optomechanical system which possesses analogous properties to an expanding universe. 
The motion of a mirror in an optical cavity induces a redshift in cavity photons. This process leads to entanglement between an effective scale factor and the sources and detectors that emulates cosmological redshift. Thus we can envision testing quantum cosmology concepts in this toy model setting.

\section{The toy FRW cosmology}
\begin{figure*}
    \centering
    \includegraphics[width=0.8\textwidth]{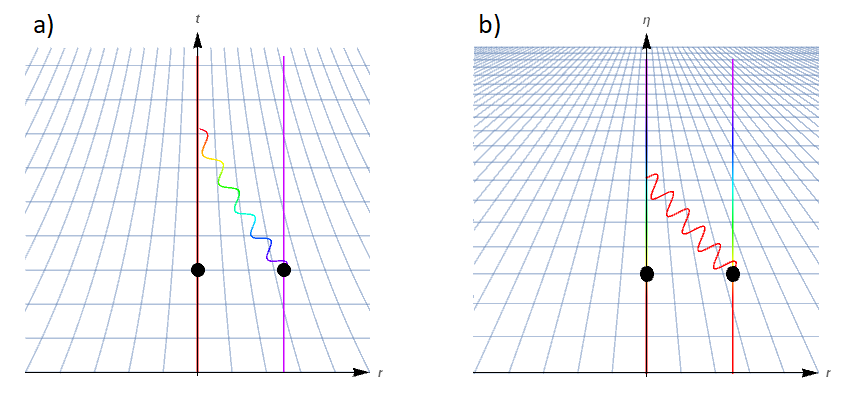}
    \caption{The FRW for an exponentially expanding universe in (a) coordinate and (b) conformal time. The universe is populated by comoving particles (black disks) which exchange a photon. Color is used to emphasize which frequencies shift.}
    \label{fig:metrics}
\end{figure*}

Consider a universe in which the metric variables are quantized. In particular, consider the Friedmann-Robertson-Walker (FRW) metric
\begin{equation}\label{eq:frwMetric}
ds^2 = -dt^2 + a^2 (t) \left( \frac{dr^2}{1-\kappa r^2} + r^2 d\Omega^2 \right)\ ,\end{equation}
where $a$ is the scale factor and $\kappa$ is the curvature. This metric is visualized in figure~\ref{fig:metrics}(a) for an exponentially expanding universe. The scale factor will be treated as a quantum variable. One then describes a cosmological wavefunction given by a superposition of eigenstates of this new $\hat{a}$ operator. This operator will then appear in various processes to account for effects of cosmological expansion.

This concept of a quantized metric relies heavily upon an underlying assumption that the various matter and energy fields `follow' the universe wavefunction, that is, that the matter fields are in a state $\ket{\Psi(a)}$ which is purely a function of $a$, and thus in the quantized theory, they follow along with changes in $a$ adiabatically. We only break this basic story for the photons, which we will allow to be excited away from this nominal matter-universe state. In a fundamental sense, this is looking at only the very lowest energy excitations: photons and gravity. We will denote this matter-simplified toy cosmology ``tFRW''.

Photons are known to be effected by the cosmological scale factor. In particular, the expanding universe results in a cosmological redshift from distant objects~\cite{hubble_relation_1929,sitter_distances_1930}. If a photon is emitted at frequency $\omega_0$ at a scale factor of $a_0$ and is later observed at a scale factor of $a_1$, its observed frequency will be $\frac{a_0}{a_1}\omega_0$. The entanglement between photons and the metric implies that measuring (the redshift of) a photon will restrict the universe's wavefunction. Then, due to constant measurements of photons in the universe, the metric variable should essentially behave classically. The manner in which information from the metric is encoded in the photon will be explored. 


In building the tFRW cosmology, we first need to choose a metric. The FRW metric (equation~\ref{eq:frwMetric}) is a simple and highly-symmetric choice, since it is homogeneous and isotropic in spatial slices (i.e., at constant time). Moreover, a single scale factor, $a(t)$, describes the dynamics of the universe. Now, a reparametrization can be applied to the metric
\begin{equation}\label{eq:frwMetricConf}
ds^2 = a^2 (t) \left[ -d\eta^2 + \left( \frac{dr^2}{1-\kappa r^2} + r^2 d\Omega^2 \right) \right]\ ,
\end{equation}
where $d\eta^2 = \left( a(t) \right)^{-2} dt^2$ describes the conformal time. This coordinate transformation is visualized in figure~\ref{fig:metrics}(b).

Consider a photon in an FRW universe. Denote the phase of the photon with $\theta$. Then, the coordinate frequency is $\omega=\frac{d\theta}{dt}$. If $a(t_0)=1$, then the coordinate frequency is given by 
\begin{equation}\label{eq:freqScaling}
\omega(t)=\frac{1}{a(t)} \frac{d\theta}{dt}\biggr|_{t_0}  = \frac{1}{a(t)} \omega_0\ ,
\end{equation}
where $\omega_0$ is a nominal frequency (taken at some $t_0$).
Moving to conformal time, the conformal frequency 
\begin{equation}
\tilde{\omega}(\eta)=\frac{d\theta}{d\eta}
\end{equation}
is then constant in time. This invariance follows from Maxwell's equations being conformal \cite{tsagas_electromagnetic_2005}. Thus, when our only degrees of freedom are photons and gravity, it is natural to build the tFRW cosmology in conformal time, leading to a simple picture for Hilbert space.

The goal now is to describe a universe in which the scale factor is treated as a quantum variable. Begin by defining a universe wavefunction in terms of the scale factor eigenbasis
\begin{equation}
\ket{u_0} = \int h_0(a) \ket{a} da\ ,
\end{equation}
where $h_0(a)$ is the universe wave function and $\ket{a}$ is an eigenvector of $\hat{a}$, satisfying
\begin{equation}
\hat{a}\ket{a} = a\ket{a}\ .
\end{equation}

Consider a simple quantum system consisting of a photon source, detector, vacuum (containing photon modes), and universe. Denote the initial state corresponding to each of these components as
\begin{equation}
\ket{e}_S\ket{g}_D\ket{\text{vac}}\ket{u_0}\ ,
\end{equation} 
where $\ket{g}$ and $\ket{e}$ refer to the ground and excited states, respectively. Unlike typical treatment of the photon mode, the photon's conformal frequency is recorded instead of its coordinate frequency. The system will undergo the following process: a photon is emitted from the source, traverses the universe for some time, and then is absorbed/measured by the detector. 

Photon emission must then be described in terms of conformal frequency. In coordinate time, the photon frequency is given by the wavepacket, $f(\omega)$. The emitted wavepacket (in coordinate frequency) is invariant of the scale factor \cite{bonnor_size_1999,nowotny_hydrogen_1972}. This is equivalent to saying that the local behavior of atoms is independent of the present cosmology. Moving into conformal frequency, $\omega\mapsto \omega/\hat{a}$, the photon's conformal frequency wavepacket generator (normalized) becomes
\begin{equation}\label{eq:emitFreq}
\hat{b}_{(\omega_0)}^\dagger = \int \frac{1}{\sqrt{\hat{a}}} f\left( \frac{\omega}{\hat{a}} \right) \hat{b}_\omega^\dagger d\omega\ ,
\end{equation}
where $\hat{b}_\omega^\dagger$ generates a single conformal frequency mode.
Similarly, the photon packet annihilator is
\begin{equation}\label{eq:absorbFreq}
\hat{b}_{(\omega_1)} = \int \frac{1}{\sqrt{\hat{a}}} g^*\left( \frac{\omega}{\hat{a}} \right) \hat{b}_\omega d\omega
\end{equation}
which annihilates the (coordinate frequency) wavepacket given by $g(\omega)$. Observe that physicality demands the scale factor be strictly positive; a negative scale factor is non-physical, and a zero scale factor corresponds to a singularity (e.g., the Big Bang). In this construction, the sources/detectors will be time-dependent (via coupling to the time-varying universe wavefunction) while the photon mode is time-independent. The wavepacket creation/annihilation operators entangle the photon frequency with the scale factor; the magnitude of a given photon mode is dependent on the state of the universe.

A few more operators are needed to fully describe the system. First, the creation operator will have a $\ketbra{g}{e}_S$ factor to reflect the source moving to its ground state with the emission of a photon. Similarly, the annihilation operator will have a $\ketbra{e}{g}_D$ factor to reflect the absorption of the photon. After detection, the system is projected into the detected state with $\hat{\Pi}=\ketbra{e}{e}_D$. Events in which no photon is detected are thrown out as inaccessible; they are inaccessible in the sense that one does not gain any information on the scale factor when no measurement occurs. Thus, the act of measuring is equivalent to post-selecting the universe in which the measured event occurred. Finally, the universe evolves according to the general operator
\begin{equation} \label{eq:FRWEvolve}
\mathcal{U} \ket{\tilde{a}} = \int B(a, \tilde{a}) \ket{a} da
\end{equation}
for some evolution function $B:\mathbb{R}^+\times\mathbb{R}^+\to\mathbb{C}$ given in the $\hat{a}$ eigenbasis. The behavior of $B$ is determined by the cosmological equations of state, which in turn could be a complicated system determined by adiabatically following matter-energy state $\ket{\Psi(a)}$. Here we will neglect the specific role the emitters and detectors play in this equation of state, appropriate for the FRW assumptions.

The system's evolution is described by
\begin{equation}
\hat{\Pi}\hat{b}_{(\omega_1)}\mathcal{U}\hat{b}_{(\omega_0)}^\dagger\ .
\end{equation}
The final state of the universe is then
\begin{equation}\label{eq:scaleMeasure}
h_1 (a) \propto \int q(a,c) B(a,c) h_0(c) dc\ .
\end{equation}
The function
\begin{equation}\label{eq:suppFactor}
q(a,c) = \frac{1}{\sqrt{ac}} \int g^*\left( \frac{\omega}{a} \right) f \left( \frac{\omega}{c} \right) d\omega
\end{equation}
describes how the source/detector wavepackets affect the final state of the universe. Equation~\ref{eq:suppFactor} can also be expressed as a function of $\frac{a}{c}$ --- this is clear when applying the change-of-variables $\omega\mapsto c\omega$. This property is expected because photon measurement yields information about the proportion between initial and final scale factors. Observe that equation~\ref{eq:scaleMeasure} demands a proportionality constant. This is to account for the non-zero probability that the photon is not detected. The not-detected state is either removed when applying $\hat{\Pi}$ or throwing out the event altogether as inaccessible via post-selection.

\subsection{Simple examples}

Consider a simple evolution in which all scale factors scale by a constant, $s$. Thus,
\begin{equation} \label{eq:simpleEvolve}
    B(a,\tilde{a}) \approx \delta(a-s\tilde{a})\ .
\end{equation}
where additional corrections necessary to maintain probability (that make $B$ unitary) are neglected for simplicity.
Solving for equation~\ref{eq:scaleMeasure}:
\begin{equation}\label{eq:simpExRes}
    h_1(a)\propto\frac{1}{a\sqrt{s}}\left( \int g^*\left( \frac{\omega}{a} \right) f \left( \frac{s \omega}{a} \right) d\omega \right) h_0\left( \frac{a}{s} \right)\ .
\end{equation}
As one would expect, if $h_0$ is centered around $a_0$, then $h_1$ is centered around $a_1 = s a_0$. Moreover, the coefficient is maximized for $g(\omega/s)=\sqrt{s}f(\omega)$. That is, if $f$ is centered around $\omega_0$, then $g$ is centered around $\omega_0/s$. This is the case when the absorption wavefunction perfectly aligns with the shifted packet so equality now holds in equation~\ref{eq:simpExRes}.

\begin{figure}
\centering
\includegraphics[width=0.8\columnwidth]{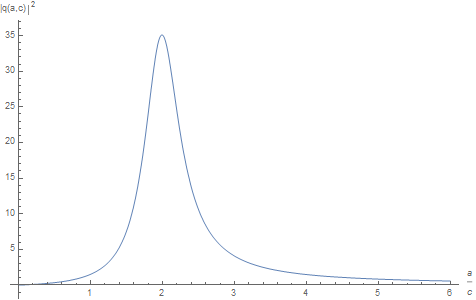}
\caption{The result of $q(a,c)$ for the simple example with the Lorentzian wavefunctions. Here, $\gamma_0=\gamma_1=\Gamma_0=\Gamma_1=c=1$, $\omega_0=10$, and $\omega_1=5$.}
\label{fig:simpExSuppFact}
\end{figure}

Now, consider a simple case for the photon wave functions. According to Weisskopf-Wigner theory on spontaneous emission, the photon wave functions should be Lorentzian~\cite{berman_spectrum_2010}: 
\[
f(\omega) = \frac{i \gamma_0^*}{\Gamma_0/2-i(\omega-\omega_0)}
\]
and
\[
g^*(\omega)=\frac{-i\gamma_1}{\Gamma_1/2+i(\omega-\omega_1)}\ .
\]
Solving for equation~\ref{eq:suppFactor} (integrating over positive and negative frequencies for simplicity) yields
\begin{equation}
q(a,c) = \frac{2\pi \gamma_0^* \gamma_1 \sqrt{a/c}}{(\Gamma_0/2 + i\omega_0) + (\Gamma_1/2-i\omega_1) a/c}\ .
\end{equation}
Figure~\ref{fig:simpExSuppFact} describes this function. The peak in this function reflects how the back-action from measuring the photon restricts the final state of the universe (i.e., metric variables); the peak is centered near the expected scale factor proportion given the photon measurement.


\subsection{As a measurement of scale factor}
One can see from equation~\ref{eq:scaleMeasure} that measuring the photon's frequency (partially) collapses the universe wave function. Consider, the extreme case where $f(\omega)=\delta(\omega-\omega_0)$ and $g(\omega)=\delta(\omega-\omega_1)$, hence a scaling of $s=\omega_0/\omega_1$. This results in
\begin{equation}
q(a,c) = \frac{1}{s\sqrt{ac}} \delta(c\omega_0-a\omega_1)
\end{equation}
and
\begin{equation}
h_1(a) \propto \frac{1}{\omega_0 a\sqrt{s}} B\left( a, \frac{a}{s} \right) h_0\left( \frac{a}{s} \right)\ ,
\end{equation}
which implies the scale factors scale by $s$ from the initial state. Observe that the final state is still affected by evolution effects, though these effects are trivial for the simple evolution (equation~\ref{eq:simpleEvolve}).

\subsection{Multiple measurements}

Consider the case in which multiple photons are emitted and later measured. In particular, say $N$~photons are detected, such that all photons are emitted before any are measured. Assume that some time passes between each emission, measurement, and between the emissions and measurements. Thus, there are $2N-1$~total periods of evolution; label these with the functions $B_i$ for $i=0,1,\ldots,2(N-1)$. Assume that the photons do not interfere (i.e., they are spatially separated) so we can define $q_j$, for $j=0,1,\ldots,N-1$, to describe the effects of measuring the $j^\text{th}$ photon; if the photons interfered, we would not be able to separate the $q_j$'s into functions that consider only one photon. The wave function of the universe after $k$~photon measurements, tracing over photon modes, is
\begin{equation}\label{eq:multiPhoton}
    h_k(a) \propto \int \prod_{i=0}^{N+k-2} dc_i\ B_i(c_{i+1},c_i) * \prod_{j=0}^{k-1} q_j(c_{j+N}, c_j)*h_0(c_0)\ ,
\end{equation}
where $c_{N+k-1}=a$. Equation~\ref{eq:multiPhoton} reduces to equation~\ref{eq:scaleMeasure} when $N=k=1$.

Assume that the universe is approximately static between emissions and between measurements. That is, the universe only evolves between the last emission and first measurement. Then, we can assume $B_i(a,c)\approx\delta_{ac}$ for $i\neq N-1$. To make an additional simplification, assume that all the detectors are the same and that all the emitters are the same, i.e., $q_j\equiv q$, independent of $j$. Thus, equation~\ref{eq:multiPhoton} becomes
\begin{equation}\label{eq:simpMultiPhoton}
    h_k(a) = \mathcal{N} \int dc\ q^k(a,c) B_{N-1}(a,c) h_0(c)\ ,
\end{equation}
for normalization constant $\mathcal{N}$. Observe that $q$, in general, describes a function that peaks around the expected scale factor. Thus, in the limit $k\to\infty$, $\mathcal{N} q^k(a,c)$ acts like a Dirac-delta function of $\frac{a}{c}$ centered around the least suppressed scale factor; i.e., the maximum of $q$. The scale factor can then be determined to arbitrary accuracy with sufficiently many measurements. 

\section{Analogy: optomechanical cavity}

Direct experimental probes of cosmological expansion are limited due to the spatial and temporal scales involved. However, it may be possible to emulate cosmological processes with accessible experimental apparatus. Such an experiment must possess the same or analogous properties of the cosmology being emulated. For example, one may want to consider a process in which photon redshift occurs: such as Doppler shift, gravitational redshift, and cosmological redshift. Here, we will consider the frequency shift induced in an optical cavity by a variable cavity length.

\subsection{Cavity as space}
Consider a Fabry-P\'{e}rot interferometer: an optical cavity consisting of two mirrors. The cavity only permits integral frequencies,
\begin{equation}
\omega_{\text{cav},n}=\frac{n\pi c}{2L}\ ,
\end{equation}
where $n$ is some positive integer (the frequency mode), $c$ is the speed of light, and $L$ is the length of the cavity. Typically, the cavity is populated by only one mode, denoted simply $\omega_\text{cav}$, when e.g., the free spectral range is large. This is because a source will populate the cavity by supplying a bandwidth of frequencies much smaller than the free spectral range of the cavity $\Delta\omega_\text{FSR}=\frac{\pi c}{2L}$ around $\omega_\text{cav}$.

If the position of one of the mirrors is classically fixed while the other is allowed to move (within some potential), the optical cavity becomes coupled to the mechanical motion of the second mirror. This is seen in $\omega_{\text{cav}}$ which is dependent on the length of the cavity. Choosing coordinates where the fixed mirror is at $x=0$, the Hamiltonian of the system becomes
\begin{align}
\nonumber
\hat{H} &= \hbar \omega_{\text{cav}} \hat{b}^\dagger \hat{b} + \hat{H}_\text{mech}(\hat{x},\hat{p}) \\
&= \hbar \frac{n\pi c}{2\hat{x}} \hat{b}_n^\dagger \hat{b}_n + \frac{\hat{p}^2}{2M} + \hat{V}_\text{mech}(\hat{x})\ ,
\end{align}
where $\hat{x}$ and $\hat{p}$ are the operators associated with the position and momentum of the mirror's mechanical motion, and $\hat{b}_n$ and $\hat{b}_n^\dagger$ are the annihilation and creation operators of the $\omega_{\text{cav},n}$ mode (subscript $n$ sometimes omitted for single-mode case). The energy associated with the mechanical system (with mass $M$) is given by $\hat{H}_\text{mech}=\frac{\hat{p}^2}{2M} + \hat{V}_\text{mech}(\hat{x})$. Observe that the singularity at $x=0$ is avoided, because this is associated with a zero-length cavity. A harmonic oscillator is often used for the mechanical system (such as a mirror on a spring) due to its symmetry with the optical modes (for a review of optomechanical cavities, see, e.g., ref.~\cite{aspelmeyer_cavity_2014}). If additional modes are present in this system, one would sum over $n$ on the first term for all possible modes. 

Motivated by the FRW universe, consider defining the scaling operator 
\begin{equation}
\hat{a}_{\text{OM}}=\frac{\hat{x}}{x_0}
\end{equation}
for an arbitrarily chosen position $x_0$. Likewise define $\hat{\pi}=x_0 \hat{p}$ to obey the canonical commutation relation, $[\hat{a}_{\text{OM}},\hat{\pi}]=i\hbar$. Now, if a physical $\hat{H}_\text{mech}$ can be chosen such that the evolution of the analogy system is consistent with that of the (FRW) universe, this system can be experimentally explored.

In the Heisenberg picture, solving for the evolution of the $\hat{a}_{\text{OM}}$ and $\hat{\pi}$ operator yields:
\begin{align}
\dot{\hat{a}}_{\text{OM}}(t) &= \frac{\hat{\pi}}{x_0^2 M}\ , \\
\dot{\hat{\pi}}(t) &= \frac{i}{\hbar} \left( \hbar [\omega_{\text{cav}}, \hat{\pi}] \hat{b}_n^\dagger \hat{b}_n + [\hat{V}_{\text{mech}}, \hat{\pi}] \right)\ ,
\end{align}
where $\hat{H}_\text{mech} = \frac{\hat{p}^2}{2M}+\hat{V}_{\text{mech}}$. The dot, $\dot{\square}$, denotes the derivative in (coordinate) time. Combining these equations, and noting the commutation relation $[f(\hat{x}),\hat{p}]=i\hbar f'(\hat{x})$, yields the equation of motion
\begin{equation}\label{eq:optMechMotion}
x_0 \ddot{\hat{a}}_{\text{OM}}(t) = \frac{\hbar n\pi c}{2 x_0^2 M} \frac{\hat{b}_n^\dagger \hat{b}_n}{\hat{a}_{\text{OM}}^2} - \frac{1}{M} \hat{V}_{\text{mech}}'(\hat{x})\ .
\end{equation}

Consider the simple case of a free mirror (i.e., $\hat{V}_{\text{mech}}=0$). Then $\ddot{\hat{a}}_{\text{OM}} = \frac{\hbar n \pi c}{2x_0^3 M} N \hat{a}_{\text{OM}}^{-2}$ for a cavity with $N$ photons --- assume $N\neq 0$ to avoid trivial case. Observe that, regardless of initial conditions, this corresponds to an (eventually) expanding cavity. After a long time, the mirror will move outward at a constant velocity. One can think of this as radiation pressure pushing the mirrors apart. As the mirrors separate, the cavity frequency decreases. Thus there is less pressure on the mirror. Eventually, the mirror will move at an approximately constant velocity, as if no longer being accelerated by the low-energy photons. Observe that the redshift and period of the photon traversing the cavity each contribute a $\hat{a}_{\text{OM}}^{-1}$ factor to the acceleration of the mirror, hence $\ddot{\hat{a}}_{\text{OM}} \propto \hat{a}_{\text{OM}}^{-2}$. As with the FRW model, the presence of radiation drives the dynamics of the system.

In this single-cavity construction, the cavity can be thought of as the space through which light travels. A source populates the cavity with a given (``coordinate'') frequency. The photons populate the cavity for some time at the same cavity mode (``conformal frequency'') then are emitted to a detector. The detector may measure a different frequency than what the source output, as the cavity frequencies change. Measuring the shift from the emitted to the detected frequency yields information on how the cavity position changed. The choice in mechanical potential determines the dynamics of the system. 

Various connections can be made between the tFRW cosmology and optomechanical system described here. The coordinate frequency (analogy) of the cavity is $\omega = \frac{n\pi c}{2 L}$, which scales correctly with the scale factor (analogy), $\hat{a}_{\text{OM}}$. The conformal frequency is analogous to the cavity mode $n$ --- a constant quantity associated with a traversing photon that is essentially inaccessible to an observer that cannot measure the mirror position (scale factor) directly. Equivalently, one can define $\omega_0 = \frac{\pi c}{2x_0}n$ as the analogous conformal frequency.



Observe that the evolution of the universe is described in terms of the optomechanical equations of motion instead of the reverse. Thus, one wants to start with a model for cosmological equations of motion then choose an optomechanical system that can mimic these equations. The optomechanical system is modified via the potential on the mechanical system, $\hat{V}_\text{mech}$. 

\subsection{Cavities as atoms}

\begin{figure}
\centering
\includegraphics[width=0.8\columnwidth]{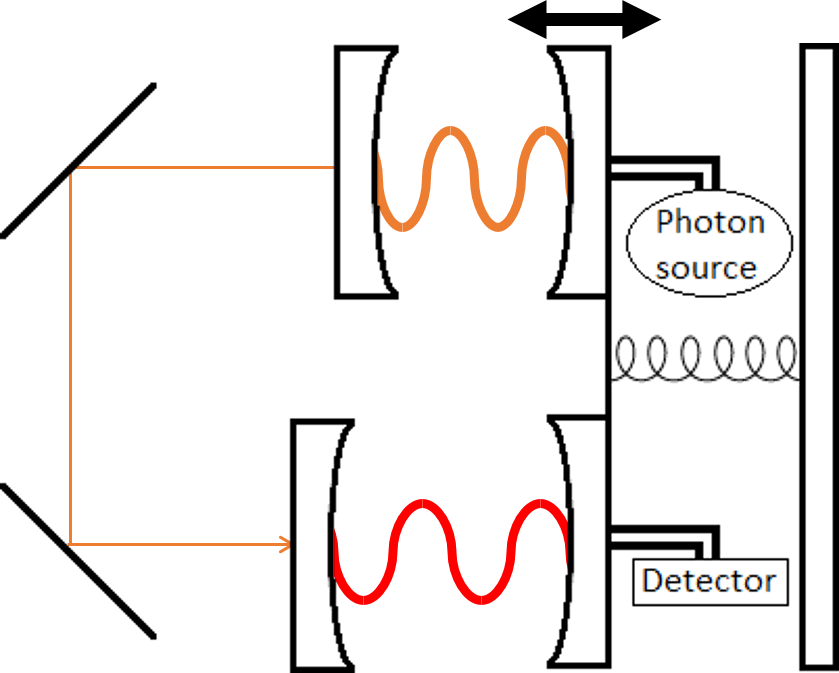}
\caption{A schematic of the coupled two-cavity system. Mirrors to the left of the cavities enable photons to transfer between cavities. The right cavity mirrors share a coupling to a mechanical system (denoted as a spring in the schematic). In addition, the top mirror is driven by a source, while the bottom mirror has an attached detector}
\label{fig:twoCavity}
\end{figure}

Consider a system with two coupled cavities (see figure~\ref{fig:twoCavity}). In this system, two cavities each share a mechanical coupling on one mirror. The other two mirrors are classically fixed, though not necessarily at the same relative position. The cavities are also allowed to transfer photon modes. In this analogy, the cavities will be shown to behave like atoms in the tFRW model.

In this optomechanical system, one cavity (say the upper cavity) can serve as a source, while the other (lower cavity) serves as the detector. The cavity frequency is analogous to the conformal frequency while the cavity mode is analogous to the coordinate frequency. Thus, the emission spectrum of the cavities is constant with respect to cavity modes (as with the coordinate frequency wavepackets in equations~\ref{eq:emitFreq} and~\ref{eq:absorbFreq}). The shared mechanical mode reflects the universe wavefunction. 

Consider a process in which a photon is emitted by the upper cavity, the cavities are allowed to shrink, then the photon is absorbed by the lower cavity. The lower cavity would see a redshifted photon in this process as the respective cavity modes increase in frequency, while the en route photon frequency is constant. This could be thought of as atoms appearing ``smaller'' as the universe expands.  We remark here that the reverse process (detector to source) is in principle allowed both in the real universe and our toy model, but we will assume no emission from the detector cavity at present. Crucially, we assume a relatively long delay (larger than the inverse linewidths of the cavities) for the photon path that exits one cavity and enters the other.

Denote the frequency in the cavities as $\omega_k=\frac{n_k\pi c}{2(\hat{x}+x_k)}$ for $k=1,2$ and $x_k$ refer to the relative positions of the fixed mirrors to the equilibrium position of the moving mirrors. As defined, $\hat{x}$ yields a position relative to the mechanical equilibrium. Considering the energy from the photons in the cavities,
\begin{equation}
\hat{H}_\text{opt} = \sum_{k=1}^2 \hbar\omega_k \hat{b}_k^\dagger\hat{b}_k\ ,
\end{equation}
where $\hat{b}_k$ are the respective annihilation operators. 

We note that this equation, as stands, will not be able to explore a wide range of redshifts. However, we can do substantially better by moving to a rotating frame, e.g., by having a weak, narrowband thermal or coherent source at a frequency $\nu_k$ coming into the source cavity. More specifically, the simplest example would be a series of excited, broad linewidth atoms near the cavity frequency in the Purcell-enhanced regime, such that photon emission occurs preferentially at the cavity frequency. A more interesting option for sources for future experiments might include having an induced chemical potential $\mu = \hbar \nu_k$ for light using parametric coupling to a bath~\cite{hafezi_chemical_2015}. On the detector side, we can consider putting a broadband photodetector on the transmitting side of the detector cavity. 

\subsection{Hubble expansion in the rotating frame}
In the cavities-as-atoms picture, we can move to a rotating frame according to the unitary transformation $\hat{\mathcal{V}}=\prod_{k=1}^2 e^{-it\nu_k\hat{b}_k^\dagger\hat{b}_k}$. Then, performing a linear expansion about small displacements of the cavity modes gives:
\begin{equation} \label{eq:optHamRot}
\hat{H}_{\text{opt}} = \sum_{k=1}^2 -\hbar (G_k \hat{x} + \Delta_k) \hat{b}_k^\dagger\hat{b}_k\ ,
\end{equation}
where $G_k=\frac{\partial \omega_k}{\partial\hat{x}}$, and $\Delta_k = \nu_k - \omega_k(0)$ is the detuning for $\hat{x} = 0$. We assume that $\Delta_k + G_k\hat x < 0$ for the entire range of dynamics that follows to maintain a semblance of physicality for our model.

We now examine the spectrum of emission and absorption, as defined in equations~\ref{eq:emitFreq} and~\ref{eq:absorbFreq}. The cavity resonance should move with scale factor according to $\omega_k(x) = \omega_k(0)/a$, where $x=0$ is chosen to reflect $a=1$. Thus, we can define the optomechanical scale factor according to
\begin{equation}
    \label{eq:anaScaleAtoms}
    \hat{a}_\text{OM} = \frac{\Delta}{\Delta + G\hat{x}}\ ,  
\end{equation}
which can be written as
\begin{equation}
    \label{eq:anaScaleInvAtoms}
    \hat{x} = \frac{\Delta}{G} \left( \frac{1}{\hat{a}_\text{OM}} - 1 \right)\ .
\end{equation}

Applying this change-of-variables to equation~\ref{eq:optHamRot} yields
\begin{equation}
    \hat{H}_{\text{opt}} = \sum_{k=1}^2 -\hbar \left[ \left( \Delta_k - \frac{G_k}{G}\Delta \right) + G_k\frac{\Delta}{G}\frac{1}{\hat{a}_\text{OM}} \right] \hat{b}_k^\dagger \hat{b}_k\ .
\end{equation}
From this hamiltonian, we would like to emulate the frequency scaling described in equation~\ref{eq:freqScaling}. This can be achieved by demanding the first term in the above hamiltonian vanish. To achieve this, the constraint $\Delta_k/G_k=\text{const}$ is applied to the system. That is, freely choose, say, $\Delta_1/G_1=\Delta/G$ and set $\nu_k=\omega_k(0)+\frac{G_k}{G}\Delta$ as the rotating frame frequencies. Now, we can set $\Delta_k/G_k \equiv \Delta/G$ to be independent of $k$. Then, the optical part of the hamiltonian is simply
\begin{equation}
\hat{H}_{\text{opt}} = \sum_{k=1}^2 \hbar \left( \frac{\tilde{\omega_k}}{\hat{a}_\text{OM}} \right) \hat{b}_k^\dagger\hat{b}_k\ ,
\end{equation}
where each cavity has, in the rotating frame, an effective `atomic' frequency 
\begin{equation}
    \tilde \omega_k = -\Delta_k\ .
\end{equation}
Connecting to the tFRW model, $\tilde\omega_k$ is analogous to the coordinate frequency wavepacket (single-valued in this scenario) in equations~\ref{eq:emitFreq} and~\ref{eq:absorbFreq}.

Consider the dynamics of this system. It is first useful to consider the Hubble parameter as defined in cosmology
\begin{equation}
    H = \frac{1}{a}\frac{\partial a}{\partial t} = \frac{1}{a^2}\frac{\partial a}{\partial \eta}\ .
\end{equation}
Thus, the conformal time derivative of $\hat{x}$ from equation~\ref{eq:anaScaleInvAtoms} (taking $\hat{a}_\text{OM}$ as the analogous scale factor) can be written
\begin{equation}\label{eq:hubbleDynamics}
    \frac{\partial \hat{x}}{\partial\eta} = \left( -\frac{\Delta}{G} \right) H\ .
\end{equation}
Observe that, in this analogy, the time is conformal time because the photon traverses between the cavities at constant frequency. Conventionally, the Hubble parameter is approximated as a constant. With this assumption, the above dynamics implies that the mirror has constant velocity, or that the mirror's motion is free. A free mirror in the system can be obtained by considering short time scales while the mirror is near equilibrium position so that the mirror has negligible acceleration.

Speaking more practically, we can now consider the following experiment: a photon is created in the source cavity, then detected at a later time at the detector cavity. If we start with a combined wavefunction of the source, detector, and mirror of the form $\ket{s}\ket{d}\ket{\psi(a)}$, the post-selected wavefunction after the detection has selected a narrow range of allowed $a$ values, according to $|q(a,c)|^2$. At the same time, the intuitive explanation --- that detection yields information about the position of the mirror --- explains the same experiment. Thus by looking for the position-based dephasing of the mirror using weak homodyne measurement of the mirror quadratures, one can estimate this `cosmological analog' decoherence.



\section{Conclusion}

A toy FRW quantum cosmology was proposed in which a metric variable was treated as a quantum variable. The scale factor in the tFRW model was promoted to the quantum operator $\hat{a}$, and a universe wavefunction was defined as a superposition of scale factor eigenstates. In this construction, it is natural to describe events in terms of conformal time, $d\eta = \frac{dt}{a}$. In particular, the photon modes were defined by their conformal frequency in lieu of their coordinate frequency. In doing so, the photon wavefunction is invariant upon time evolution, while a process that generates/annihilates a photon gains a time-dependence via a coupling to the time-dependent scale factor. 

A simple process in which a photon is emitted then measured is described. The act of measurement restricts the possible values of the universe scale factor such that the measured redshift agrees with the change in scale factor. The universe becomes entangled with the photon resulting in a wavefunction collapse upon measurement. As photons are constantly being measured throughout the universe, one would expect the scale factor to behave classically --- in agreement with observation.

Motivated by the toy cosmology, two optomechanical analogies were proposed to emulate the cosmology. In these analogies, a mechanical system serves as the expanding universe. Optical modes couple to the mechanical system and experience frequency shifts analogous to cosmological redshift. 

The two analogies approach cosmological scaling in different ways. In the analogy in which the cavity was treated as the space through which photons traveled, the dynamics of the system are most immediately described in terms of coordinate time. That is, the en route photons' frequency shifts directly as the mechanical system evolves. In contrast, when treating cavities as atoms, the dynamics are described in terms of conformal time, because the en route photons remain static while the cavity modes shift. In either case, there is some freedom to adjust the system to reflect various cosmological dynamics (e.g., consider equations~\ref{eq:optMechMotion} and~\ref{eq:hubbleDynamics}). With further considerations, these optomechanical systems may be used to provide a testbed for experiments in quantum cosmology. 

\section{Acknowledgements}

We thank G. Milburn, S. Jordan for helpful discussions. 
Funding comes from the NSF funded MRSEC at Princeton University and the NSF funded Physics Frontier Center at the JQI. 

\bibliographystyle{apsrev}
\bibliography{qScaleBib}

\end{document}